\documentclass[%
 aip,
 amsmath,amssymb,
 reprint,%
 floatfix,
]{revtex4-1}

\usepackage{graphicx}
\usepackage{dcolumn}
\usepackage{bm}

\usepackage[utf8]{inputenc}
\usepackage[T1]{fontenc}
\usepackage{mathptmx}

\begin{document}

\preprint{}

\title{Improving resolution-sensitivity trade off in sub-shot noise imaging}

\author{I. Ruo Berchera} 
\email{i.ruoberchera@inrim.it.}

\author{A. Meda}
\affiliation{ Istituto Nazionale di Ricerca Metrologica,Strada delle Cacce 91, Torino 10135, Italy}

\author{E. Losero}
\affiliation{ Istituto Nazionale di Ricerca Metrologica,Strada delle Cacce 91, Torino 10135, Italy}

\author{A. Avella}
\affiliation{ Istituto Nazionale di Ricerca Metrologica,Strada delle Cacce 91, Torino 10135, Italy}

\author{N. Samantaray}
\affiliation{Quantum Engineering Technology Labs, H. H. Wills Physics Laboratory and Department
of Electrical and Electronic Engineering, University of Bristol, BS8 1FD, UK}%

\author{M. Genovese}
\affiliation{ Istituto Nazionale di Ricerca Metrologica,Strada delle Cacce 91, Torino 10135, Italy}

\date{\today}

\begin{abstract}
One of the challenges of quantum technologies is realising the quantum advantage, predicted for ideal systems, in real applications, which have to cope with decoherence and inefficiencies. In quantum metrology, sub-shot-noise imaging (SSNI) and sensing methods can provide genuine quantum enhancement in realistic situations. However, wide field SSNI schemes realized so far suffer a trade-off between the resolution and the sensitivity gain over classical counterpart: small pixels or integrating area, are necessary to achieve high imaging resolution, but larger pixels allow a better detection efficiency of quantum correlations, which means a larger quantum advantage. Here we show how the SSNI protocol can be optimized to significantly improve the resolution without giving up the quantum advantage in the sensitivity. We show a linear resolution improvement (up to a factor 3) with respect to the simple protocol used in previous demonstrations. 
\end{abstract}

\maketitle


\section{\label{Introduction}Introduction\\}


Imaging delicate systems using small number of incident photons with true and significant quantum enhanced sensitivity is extraordinarily important for applications, from biology and medicine to fundamental physics research. The first proof of principle of SSNI of a 2D absorption/transmission mask was given in 2010 \cite{Brida2010} and in 2017 we reported the realization of the first wide field sub shot noise microscope \cite{Samantaray2017}. It is based on spatially multi-mode non-classical photon number correlations of two-mode squeezed vacuum (TMSV) states, produced by Spontaneous parametric down conversion (SPDC) and detected by a high quantum efficiency CCD camera \cite{jedrkiewicz2004,brida2009,blanchet2008,gatti2008, Genovese16}. The sample (2D mask) is probed by one beam with certain level of quantum fluctuations, so that the detected image is affected by a noise pattern. However, a second reference beam, locally correlated in the photon number with the first one, generates at the detector an identical noise pattern. In this way, one can simply remove the noise by subtraction. The microscope of Ref. \cite{Samantaray2017} produces real-time images of several thousands pixels and $5\mu$m of resolution even though the actual quantum enhancement in sensitivity compared with the best classical protocol is effective at larger spatial scales. In fact, in this technique there is a clear trade-off between the resolution and the quantum enhancement, due to the fact that pixels or integrating areas smaller than the characteristic size of the spatial modes do not intercept all the correlated photons between pairs of conjugated modes. 
With the aim of improving the performance of this technique, in Ref. \cite{Losero18} we studied in deep detail the problem of absorption estimation by photon counting towards the ultimate quantum limit, taking into account experimental inefficiencies. In particular, we analytically showed the advantage of the optimized estimator proposed in \cite{Moreau2017}, especially in case of limited detection efficiency. This estimator does not involve modification of the set-up, but only a slightly different use of the data and a pre-calibration of the system.

Here we use this estimation protocol to shift the resolution-sensitivity trade-off of the SSNI, improving the resolution of a factor 3 in the best case. It turns out that in this way it is in principle possible to obtain SSNI at the Rayleigh resolution limit.

\section{\label{Estimation of a loss with TMSV state}Estimation of a loss with TMSV state\\}

The lower bound to the uncertainty in a loss estimation for classical probes, i.e. mixture of coherent states, is \cite{pirandola2017} 

\begin{equation}\label{Ucoh}
U_{coh}\backsimeq[(1-\alpha)/\langle n_{P} \rangle]^{1/2}
\end{equation}

where $\langle n_{P} \rangle$ is the mean number of photons of the probe and $0\leq\alpha\leq1$ is the loss induced by the the sample. Only for high losses the uncertainty can be arbitrary small, while in case of a faint loss, one retrieves the expression $U_{snl}= \langle n_{P} \rangle^{-1/2}$, usually referred as to "shot-noise-limit" (SNL). 

In general, excluding adaptive  strategies where the limit is still unknown \cite{pirandola2017}, the ultimate quantum limit (UQL) of sensitivity for a single mode probe is  $U_{uql}\backsimeq\sqrt{\alpha}\, U_{coh} $\cite{paris2007,adesso2009}, which scales much more favorably than the classical bound for small losses.
Several quantum states have been demonstrated to reach in principle this ultimate limit: single mode squeezed vacuum, with detection strategy based on photon counting and Gaussian operations, for small losses and small number of photons \cite{paris2007}; Fock states $|n\rangle$,  with photon counting, unconditionally for any $\alpha$ but if $\langle n_{P} \rangle \geq 1$ \cite{adesso2009}; TMSV state with photon counting \cite{Nair18}, unconditionally for any loss and all energy regimes \cite{illuminati}.
TMSV being the photon number entangles state:

\begin{equation}
\vert TMSV\rangle_{1,2}=\sum_{n} c_{n}\vert n\rangle_{1}\vert n\rangle_{2},
\label{SPDC_state}
\end{equation}

where the subscripts "1" and "2" represent two correlated modes, and the probability amplitude is $c_{n } \propto \sqrt{\mu^n/(\mu + 1)^{n+1}}$, $\mu$ being the mean number of photons per mode. 

\par
On the experimental side, a seminal proposal on absorption measurement with photon pairs produced by SPDC, i.e. using a faint TMSV state, was given already in 1986 \cite{Jakeman86} and a sub-shot-noise measurement of modulated absorption using SPDC has been realized few years later\cite{Tapster1991}. More recently, quantum enhanced absorption measurements have been performed by post-selected heralded single photon Fock states \cite{Whittaker2017} and also trough an active feed-forward driven by an optical shutter \cite{sabines2017}. In those cases, on/off single photon detectors have been used. However, the higher genuine quantum enhancement has been achieved in experiments exploiting low noise intensity measurement (photon counting), taking advantage of the high quantum efficiency and small electronic noise of the modern CCD cameras \cite{Moreau2017,Losero18}. An enhancement of the order of 50\% respect to the classical bound has been achieved for the same number of detected photons and 32\% if perfect detection efficiency is considered only for the classical scheme \cite{Losero18}. With these detectors, that provide also flexible spatial resolution, and exploiting the spatially multimode emission of traveling wave SPDC it has been possible to devise \cite{gatti2008}, and realize \cite{Brida2010,Brida11,Samantaray2017} wide field SSNI schemes where a 2D amplitude mask is recovered by parallel multi-parameter absorption/transmission estimation.

Ref. \cite{Losero18}, reports a systematic study of the performance achieved by several possible estimation strategies based on the detected number of photons jointly measured in the probe, $N'_{P}$, and reference $N_{R}$. Summarizing, three estimators have been considered there:

\begin{itemize}
	\item \textit{ Ratio}, as used for example in \cite{Jakeman86}
	\begin{equation} \label{S}
	S_{\alpha}= 1-\gamma \frac{N'_{P}}{N_{R}},	
	\end{equation}
	\item\textit{Subtraction}, considered for SSNI \cite{gatti2008,Brida2010,Brida2010,Samantaray2017} 
	\begin{equation}\label{S"}
	S''_\alpha= \frac{N_R-\gamma N'_P}{\langle N_R\rangle},
	\end{equation} 
	\item\textit{Optimized subtraction}, considered in \cite{Tapster1991} and \cite{Moreau2017}
	\begin{equation}\label{S'}
	S'_\alpha = 1 - \frac{N'_P - k_{opt} \Delta N_R} {\langle N_P \rangle}.
	\end{equation}
\end{itemize}

The factor $\gamma=\langle N_{R}\rangle/ \langle N_{P}\rangle$ is introduced to account for unbalancing between the mean energy detected in the probe and reference arm without the sample. It can be evaluated in a  pre-calibration of the apparatus, that should last long enough to provide an accurate determination of $\gamma$. In the third estimator, the factor $k_{opt}$ must be optimized in function of the physical parameters of the system. In particular it turns out that  $k_{opt}$ is a function of the detection efficiencies of the channels and the local excess noise. Clearly, each of the three estimation strategies is based on the idea that the common photon number fluctuations of the probe and reference beam can be suppressed or at least mitigated by a direct comparison. However, in terms of the uncertainty they behave differently. For the general expressions the reader should refer to Ref. \cite{Losero18}. For simplicity, here we consider the same detection efficiency $\eta_d$ in the two arms, i.e. $\gamma=1$. Moreover, we consider a large number $M\gg1$ of spatio-temporal realization of TMSV states (here collectively named twin-beam state), detected by each pixel in the measurement time, and mean number of photons per mode $\mu=\langle N_{P}\rangle/M\ll1$. The last constraints allow considering each pixel with Poissonian  photon number distribution and to be independent from the others in the same arm. In this case one has:

\begin{itemize}
	\item Uncertainty of the \textit{ Ratio }\\
    \begin{equation} \label{US}
    \Delta^2 S_{\alpha}\backsimeq \frac{U_{uql}^{2}}{ \eta_d}+ 2\frac{(1-\alpha)^{2}}{ \langle N_P \rangle} (1-\eta).
    \end{equation}
	\item Uncertainty of the \textit{Subtraction}\\
	\begin{equation}\label{US"}
	\Delta^{2} S''_{\alpha} = \frac{U_{uql}^{2}}{\eta_d}+\frac{2(1-\alpha)(1-\eta)+\alpha^{2}}{ \langle N_P \rangle}.
	\end{equation}
	\item Uncertainty of the \textit{Optimized subtraction}\\
	\begin{equation} \label{US'}
	\Delta^2 S'^{(TWB)}_{ \alpha, \eta} =\frac{U_{uql}^{2}}{ \eta_d}+\frac{(1-\alpha)^{2} }{ \langle N_P \rangle}  \left(1-\eta^2\right).
	\end{equation}
\end{itemize}

In the equations above, the parameter $\eta$ ($0< \eta<1$) is related to the noise reduction factor $NRF=\mathrm{Var}(N_P-N_R)/\langle N_P+N_R\rangle$ \cite{agafonov2011,bondani2007,iskhakov2016} by the relation  $NRF=1-\eta$ . The \emph{NRF} represents the level of correlation of the joint detected photon number distributions, and  can be estimated experimentally. For $0 \leq NRF<1$, the correlations are non-classical. Therefore, $\eta$ can be interpreted as the efficiency in detecting correlated photons, i.e. the probability that for a photon detected in a certain pixel in the probe arm, its twin photon is detected in the correlated pixel in the reference arm. Thus, it can be written as the product of the channel detection efficiency and a collection efficiency term, $\eta=\eta_d\cdot \eta_c$. The collection efficiency $\eta_c$ takes into account for the fact that in real systems correlated modes cannot be always perfectly detected.
In the ideal situation, assuming $\eta= \eta_d =1$, both the \textit{Ratio} in Eq. \ref{S} and the \textit{Optimized} estimator in Eq. \ref{S'} reach the UQL,  while the \textit{Subtraction} estimator in Eq. \ref{S"} approaches the UQL only asymptotically for small value of the loss $\alpha$.
However, another significant difference appears in the non ideal detection case, because of the different dependence of Eq.s \ref{US}-\ref{US'} from $\eta$. In particular, for the \textit{Ratio} and the \textit{Subtraction} estimators, the positive additive term exceeding the UQL is $\propto2(1-\eta)$, which is larger than the one for the \textit{Optimized} case $\propto(1-\eta^{2})$, for any value of $\eta$. This means that the \textit{Optimized} estimator works always better than the others, and that this advantage is larger for low efficiency $\eta$. For example, rewriting the Eq.s \ref{US}-\ref{US'} in terms of the classical bound $U_{coh}$ of Eq. \ref{Ucoh},  it is easy to see that the quantum advantage for the \textit{Ratio} or the \textit{Subtraction} estimators starts from $\eta\ge0.5$. In contrast, the twin beam state together with the \textit{Optimized subtraction} protocol performs always better than the classical bound.\\
In the next section we will show how this feature of the \textit{Optimized} estimator is particularly suited for the SSNI improvement also in terms of resolution. 

\begin{figure}[htp]
	\centering
	\includegraphics[trim={3cm 1cm 0.3cm 0cm}, clip=true, width=1\columnwidth, angle=0 ]{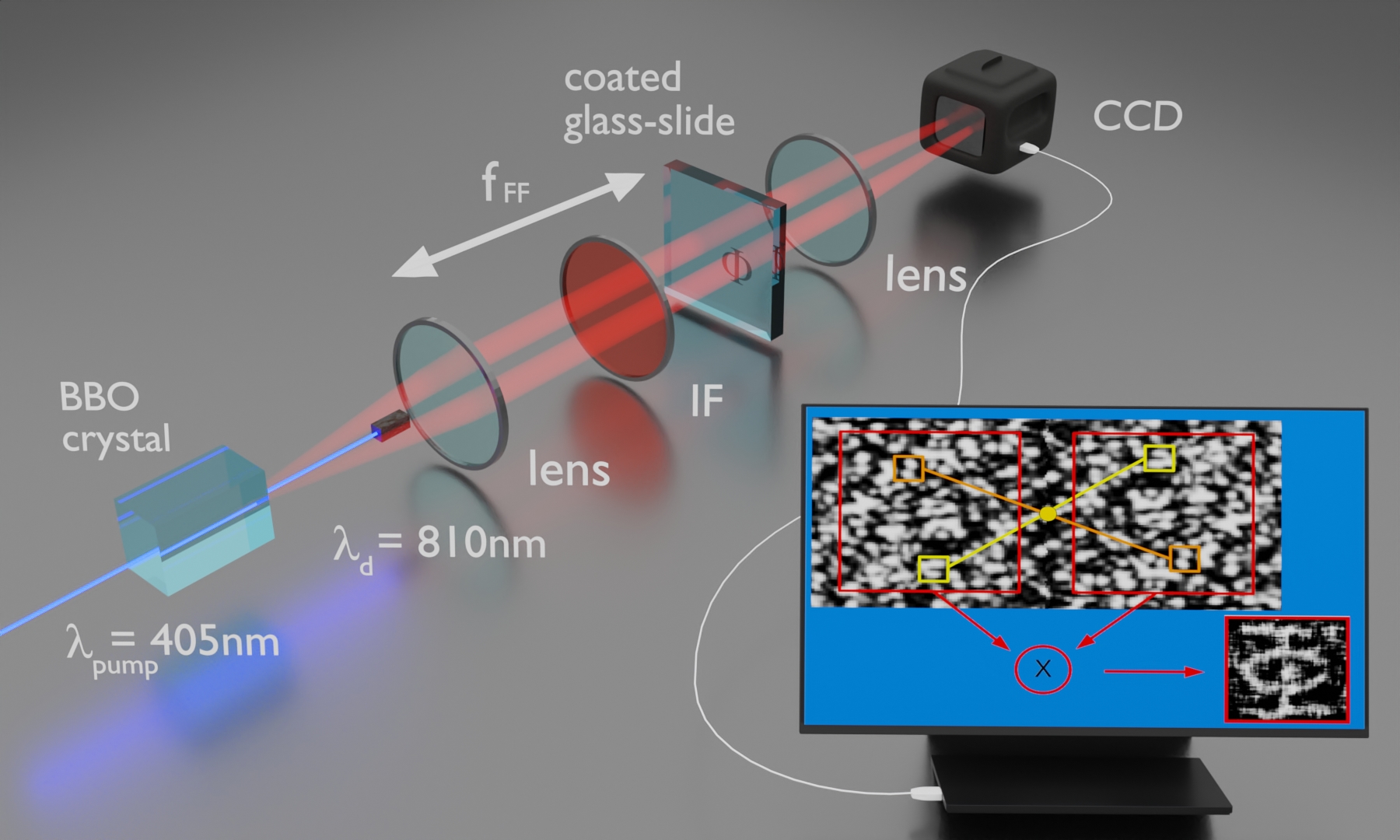}
	\caption{A multi-mode twin-beam state is produced through the SPDC, pumping a non linear crystal (Type-II-Beta-Barium-Borate, BBO) with a CW laser-beam ($100$mW at $\lambda_{pump}=405$nm). The down-converted photons around the degeneracy wavelength, $\lambda_d = 810$nm, are spectrally selected using an interferential filter (IF, $(800\pm 20$nm)). The resulting state can be approximated as a tensor product of independent TMSV states as $|\Psi\rangle = \otimes_{q,\lambda} |TMSV\rangle$, where q and $\lambda$ are the transverse momentum and the wavelength of one of the two photons produced, while momentum and wavelength of the other photon are fixed by energy and momentum conservation. The far field is obtained at the focal plane of a lens with $f_{FF} = 1$cm focal length, where the correlation in momentum is converted into correlation between symmetric positions. A coated glass-slide with a 2D absorbing mask, realized as a thin titanium deposition of absorption $\alpha \sim 1\%$, is placed in this plane. This is then imaged (magnification of 7.8) to the chip of a charged-coupled-device (CCD) camera with nominal quantum efficiency of $95\%$ at $810$nm and pixel size of $13\mu$m. We perform a $3\times 3$ pixel binning, to set the resolution to 5$\mu$m at the object plane, which matches the measured cross-correlation length. The acquisition time of a single frame is $ \sim 100$ms, the number of temporal modes  per pixel per frame $\sim10^{11}$, and the number of photo-counts $\langle N \rangle =10^3$. The estimated final detection efficiency is $\eta_{d}=0.81$}\label{setup}
\end{figure}

\section{SSNI: Experimental Results}\label{Experimental Result}

In wide field imaging realizations  with SPDC \cite{Brida2010,Samantaray2017}, the spatial pattern in the far field of the emission, where the transmitting mask is placed, is a continuous distribution of independent spatial modes with a certain coherence area given by the Fourier transform of the pump beam profile. This plane is then projected at the pixel's matrix of the detector chip, where probe and reference beams are detected in two different regions.  Fig. \ref{setup} describes the details of our experimental set-up. 

The pixel size, or more in general the elementary integration area in one arm, should be large enough to collect a certain number of modes. It is straightforward that if a photon is detected in a certain pixel in the probe arm, the corresponding pixel in the reference arm should be at least as large as the correlation area, otherwise correlated photons would fall outside the pixel, representing an effective loss when pixel to pixel correlations are considered. 
Moreover, a photon detected close to the edge between a pixel and its neighbors, has its twin photon detected with only <50\% probability in the symmetric pixel in the reference arm. Both these contributions to losses are taken into account by $\eta_c$, enclosed in $\eta$  in Eq.s \ref{US}-\ref{US'}. In the conditions of our experiment, and assuming a perfect alignment,  $\eta_c$ is solely related to the ratio $X=d/2r$, being $d$ the pixel size  and $r$ the transverse correlation radius.  Details of this model can be found in previous literature \cite{Samantaray2017,Meda14,Meda17}. 

\begin{figure}[htp]
	\includegraphics[trim={3cm 1cm 0.3cm 0cm}, clip, width=1.1\columnwidth]{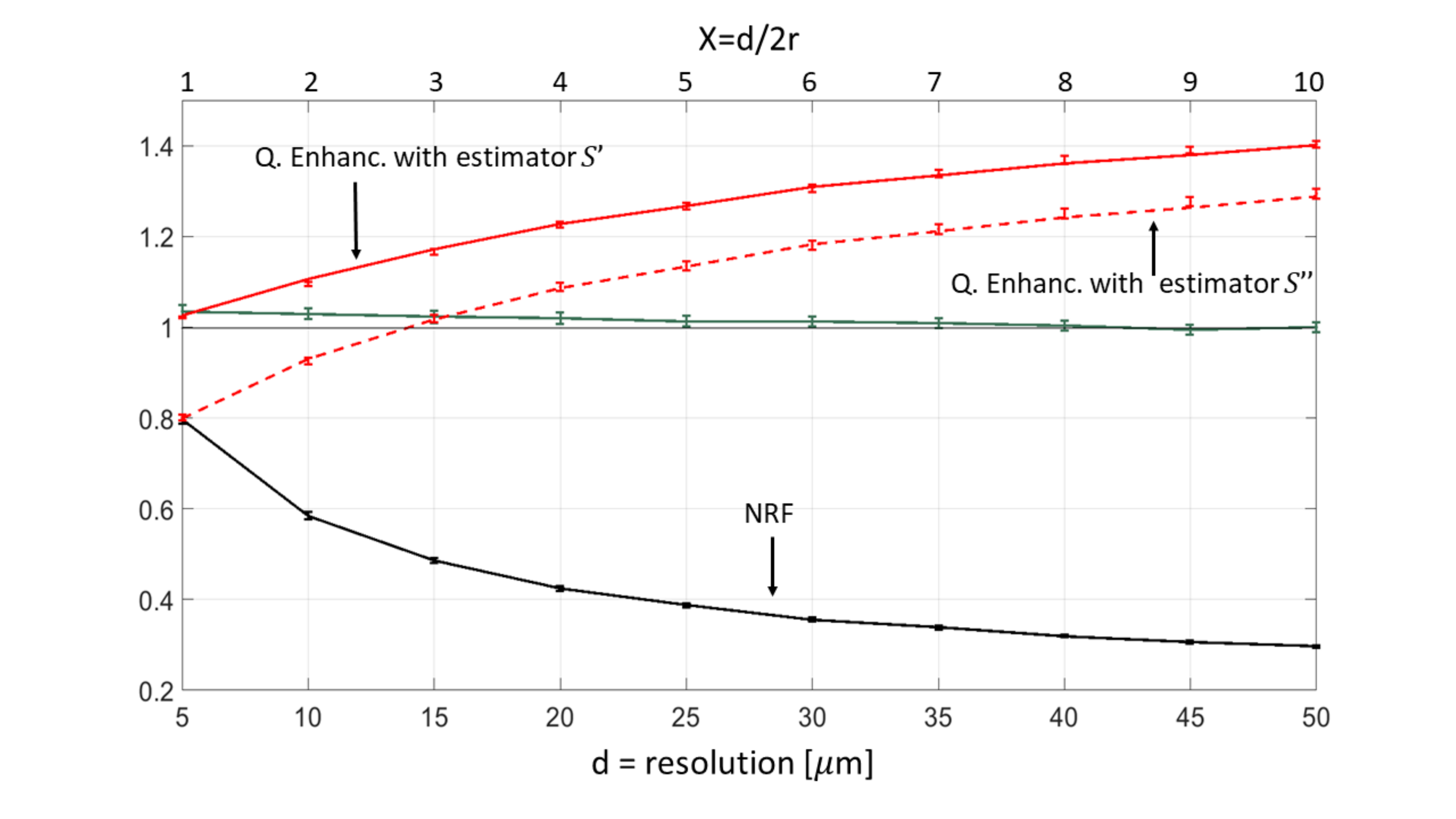}
	\caption{Experimental \emph{NRF} and quantum enhancement in function of the resolution in the object plane, $d$. The \emph{NRF} (black data-series) is evaluated as $NRF=\mathrm{Var}(N_P-N_R)/\langle N_P+N_R\rangle$ for photons numbers detected  in an area of size $d^{2}$.  Red data-series show the quantum enhancement provided by twin-beam (multimode TMSV states), both using the \emph{Subtraction} estimation strategy, in Eq. \ref{S"} (dots, dashed line), or the \emph{Optimized} one, in Eq. \ref{S'} (squares, solid line). The green data represents the quantity $\mathrm{Var}(N_P)/\langle N_P\rangle$ and confirms that the statistics of photon counts is Poissonian.}\label{QuEnhanc} 
\end{figure}

In Fig. \ref{QuEnhanc} we report the experimental $NRF$ in function of the resolution, i.e. the size of the integration area $d$. It decreases with $d$, as long as the collection efficiency $\eta_c$ increases, saturating the value $1-\eta_d$ for $X>50$ (where $\eta_c\approx1$). In the same figure also the quantum enhancement in the sensitivity is reported, in function of the resolution.  Of course, in general, a suitable trade-off between the resolution $d$, and the sensitivity should be found. The dashed red curve represents the quantum advantage of the twin beam using the \emph{Subtraction} estimator. It is evaluated as $U_{coh}/ \sqrt{\Delta^{2} S^{''}_{\alpha}}$, replacing in the Eq.s \ref{Ucoh} and \ref{US"} the values of $NRF=1-\eta$ and $\alpha$ with their experimentally estimated values. Solid red line represents the corresponding quantum enhancement for the \emph{Optimized} estimation. The data-points represent, for each case, the quantum advantage estimated by the experimental  frame-to-frame fluctuation in the absorption $\alpha$ determination, according to Eq.s \ref{S"}-\ref{S'} respectively. 300 shots and region where $\alpha\approx0.01$ are used. The experimental classical uncertainty to compare with, is obtained by the fluctuation of the estimate in  Eq. \ref{S}, where the reference $N_R$ is substituted by the mean value of the probe in absence of the sample $ \langle N_P\rangle $. This estimator, using only the probe beam, reaches the lower classical bound $U_{coh}$, so it represents the best classical strategy \cite{Losero18}. 

As we have anticipated at the end of Sec.  \ref{Estimation of a loss with TMSV state}, the quantum advantage when using the \emph{Subtraction}, as done in previous demonstration \cite{Samantaray2017}, is present for $\eta>0.5$ (red dashed line in Fig. \ref{QuEnhanc}). It corresponds to a resolution of 3 times the correlation length, namely 15 $\mu$m. We can conclude that, with this estimator it is not possible, even in principle, to have quantum enhancement and a resolution close to a single coherence length, at the same time. 

In this context, the \emph{Optimized} estimator is a big opportunity because its quantum advantage can be found also for smaller value of the efficiency $\eta$ or equivalently for $NRF\le1$. In fact, the solid line in Fig. \ref{QuEnhanc} shows that the quantum advantage is present also for $d=5\mu$m which is exactly the coherence length. Moreover this estimator performs better than the other one for any resolution, always representing the best choice for SSNI in wide field modality. 

As mentioned, the only point that deserves attention when using the optimized estimator, is that it requires a careful characterization of the experimental setup, in order to provide a reliable value of the parameter $k_{opt}$ to insert in Eq. \ref{S'}. This $k_{opt}$ is a simple function of the excess noise end the detection efficiencies in both channels. We estimated the absolute quantum efficiency with a method that can be applied with an identical setup configuration\cite{Meda14}. We found that the performance of the \emph{Optimized} estimator is not dramatically affected by the accuracy in the parameter's determination: few percent of uncertainty is enough to recover the advantage predicted by the theory.

\begin{figure}[ht]
	\includegraphics[trim={0.3cm 0cm 0.3cm 0cm}, clip, width=\columnwidth]{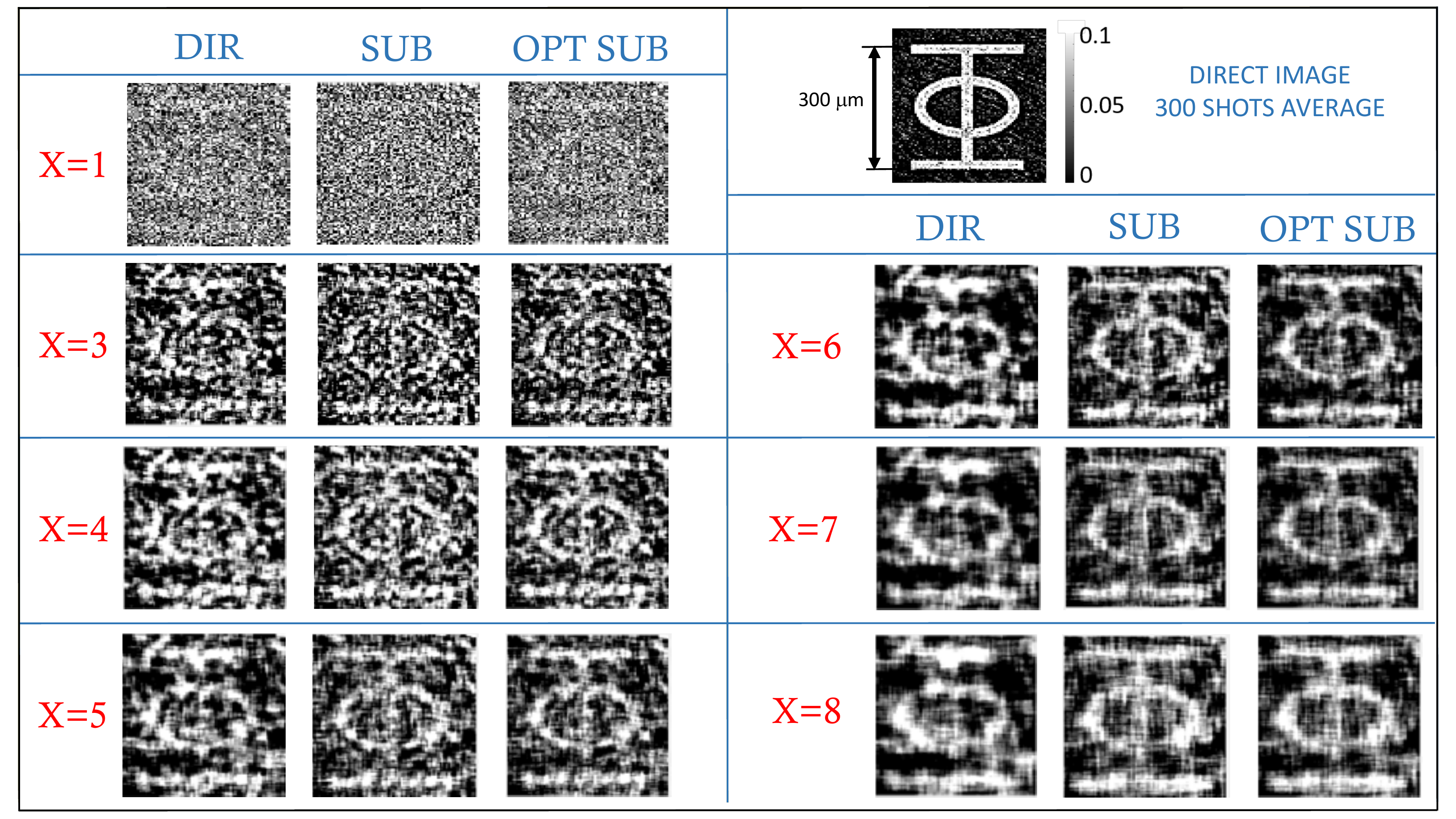}
	\caption{Comparison between single shot images for different ratios $X$ between the pixel dimension and the correlation diameter. For each $X$,  the direct (DIR) image is compared to the image obtained with the quantum \emph{Subtraction} protocol(SUB) and the one with the \emph{Optimized} protocol (OPT SUB). In the upper-right panel the direct image of the object averaged over 300 shots is reported.}\label{Phi_d} 
\end{figure}

Finally, in Fig.\ref{Phi_d} we present a single frame experimental image of a specific absorbing mask for different spatial resolutions. The mask is realized by a thin "$\mathrm{\Phi}$-shaped"    metallic deposition on a coated glass-slide with $\alpha \sim 1\%$. The resolution is set by the application of a median filter, which substitutes in each pixel (corresponding to $5\mu$m in the object plane) the mean photon counts over a square of side $d$, centered in the pixel. As expected, the images obtained with the quantum protocol, i.e. using the twin-beam state, are visually better than the ones obtained by single beam classical approach. Moreover, one can appreciate an improvement of the \emph{Optimized} estimation protocol with respect to the \emph{Subtraction} protocol in the residual noise level.

\section{Conclusion}\label{Conclusion} 
In this letter, we have shown a substantial improvement of the performance of the SSNI technique with respect to previous realizations \cite{Brida2010,Samantaray2017, sabines19}. By studying different pure loss estimations strategies with quantum states of light in presence of imperfections we demonstrate that the robustness of an \emph{Optimized} estimator with respect to detection losses, and the link between spatial resolution and inefficiencies in detecting correlated photons, implies that such estimator produces a significant advantage also in terms of resolution. 
We have demonstrated that, differently from the previous \emph{Subtraction} protocol, the limit to the resolution is given by the coherence area of the correlation in the far field of the SPDC process, that can be in principle reduced down to the Rayleigh limit determined by the numerical aperture of the optical system.
This result represents a further step toward practical applications of quantum correlations in imaging.

\begin{acknowledgments}
This work has been supported by EMPIR 17FUN01 ‘BeCOME’, the EMPIR initiative is co-funded by the EU H2020 and
the EMPIR Participating States, and by the Horizon
2020 research and innovation program under grant agreement number 862644 (FETopen-
QUARTET).

\end{acknowledgments}

{}




\end{document}